\documentclass{article}
\pdfoutput=1
\usepackage{jheppub}
\usepackage{amsmath,hyperref,amssymb}
\usepackage{graphicx}
\usepackage{dcolumn}

\usepackage{siunitx}
\usepackage{cleveref}
\usepackage{multirow}

\usepackage{soul}
\usepackage{bm}
\usepackage{verbatim}
\usepackage{tabularx}
\usepackage{slashed}
\usepackage{cancel}
\usepackage[boxsize=2em,aligntableaux=center]{ytableau}
\usepackage{float}

\usepackage{comment}
\excludecomment{toexclude}
\includecomment{toinclude}

\def\be{\begin{equation}}
\def\ee{\end{equation}}
\def\bea{\begin{eqnarray}}
\def\eea{\end{eqnarray}}

\newcommand{\eV}{\,{\rm eV}}

\newcommand{\MeV}{\,{\rm MeV}}
\newcommand{\GeV}{\,{\rm GeV}}

\definecolor{seagreen}{rgb}{0.180392,0.545098,0.341176}

\begin{document}

\title{Opening up the QCD axion window}
\author[a]{Prateek Agrawal,}
\affiliation[a]{Harvard University,
Department of Physics, Harvard University, Cambridge, MA 02138, USA}
\author[b]{Gustavo Marques-Tavares}
\affiliation[b]{Stanford Institute for Theoretical Physics, Stanford
University, Stanford, CA 94305, USA}

\author[c]{and Wei Xue}
\affiliation[c]{Center for Theoretical Physics, Massachusetts
Institute of Technology, Cambridge, MA 02139, USA}

\abstract{ 
  
  We present a new mechanism to deplete the energy density of
  the QCD axion, making decay constants as high as $f_a \simeq
  10^{17}\GeV$ viable for generic initial conditions.  In our setup,
  the axion couples to a massless dark photon with a coupling that is
  moderately
  stronger than the axion coupling to gluons. Dark photons are produced
  copiously through a tachyonic instability when the axion field starts
  oscillating, and an exponential suppression of the axion density can
  be achieved.  
  For a large part of the parameter space this
  dark radiation component of the universe can be observable in
  upcoming CMB experiments. 
  Such dynamical depletion of the axion density ameliorates the
  isocurvature bound on the scale of inflation.
  The depletion also amplifies the power
  spectrum at scales that enter the horizon before particle
  production begins, potentially leading to axion
  miniclusters.  
}

\maketitle

\section{Introduction}

Dark matter remains the most compelling evidence of physics beyond the
Standard Model~(SM). Since we have only observed it through its
gravitational effects, there is a wide range of possible dark matter
candidates. A very interesting possibility is that dark matter is made
of light bosonic fields which only interact extremely
weakly with the SM~\cite{Arias:2012az}. 
The QCD axion
is a particularly well-motivated
example of such
weakly-coupled light 
fields~\cite{Peccei:1977np,Peccei:1977hh,Weinberg:1977ma,Wilczek:1977pj} 
(see~\cite{Olive:2016xmw} for a review). 
An approximate shift symmetry protects
its mass and couplings, so that it can be naturally light and have
a lifetime much longer than the age of the universe. 
An axion field oscillating about the minimum of its potential has the
equation of state of matter, making it a dark matter candidate~\cite{Abbott:1982af,Dine:1982ah,Preskill:1982cy}. 
Further, the
axion makes the CP-violating QCD $\theta$ angle dynamical, with a
minimum at
$\bar{\theta}=0$~\cite{Vafa:1984xg}. 
Thus, the axion can naturally explain the severe constraints on CP
violation in the strong sector, $\bar{\theta}\lesssim 10^{-10}$,
from null measurements of the neutron/Hg
electric dipole moments~\cite{Baker:2006ts,2016PhRvL.116p1601G,Afach:2015sja}.

The mass of the QCD axion is
set by its
decay constant $f_a$. Its couplings to other SM fields are
model-dependent, but are expected to be
given by higher-dimension operators suppressed by 
$f_a$.
Astrophysical observations impose a lower bound on the decay
constant, $f_a\gtrsim 10^{9}\GeV$ for generic
couplings~\cite{Andriamonje:2007ew,Ayala:2014pea,Abramowski:2013oea,TheFermi-LAT:2016zue,Payez:2014xsa}
(for detailed discussion of the constraints see~\cite{Olive:2016xmw}). 
On the other hand,
axion decay constants of $f_a>10^{12}\GeV$ predict a dark matter
abundance in excess of observations
for generic initial conditions.
These two considerations determine the axion
window, $10^9\GeV< f_a <10^{12}\GeV$, 
within which axions can naturally account for the observed
dark matter density. 
There are a number of ongoing and proposed experiments for probing the
QCD axion window~\cite{Asztalos:2010,Shokair:2014rna,
Armengaud:2014gea,Arvanitaki:2014dfa,Brubaker:2016ktl}.

We propose a new mechanism for depleting the axion
abundance in the early Universe, such that the region
$f_a>10^{12}\GeV$ becomes viable. This region is well-motivated
theoretically.
Generic expectations from string theory predict
$f_a$ values higher than the axion
window~\cite{Banks:2003sx,Svrcek:2006yi,Conlon:2006tq}. 
Beyond theoretical arguments, a
practical reason for exploring viable cosmological models
 outside the axion window is that there are a number of exciting
proposed experiments that should have sensitivity to very light (and thus large-$f_a$) axions ~\cite{Budker:2013hfa,Arvanitaki:2014wva,
Silva-Feaver:2016qhh,Kahn:2016aff,Arvanitaki:2016qwi,Garcon:2017ixh}. 

Our mechanism uses particle production in the time-dependent axion
background. 
We assume that the axion couples to a massless dark photon, a $U(1)$ gauge
boson that is decoupled from the SM.
As the axion starts oscillating in the early universe, 
certain modes of the gauge field become tachyonic, leading to an 
exponential growth of such modes.
A similar mechanism has been previously used for generating primordial
magnetic fields~\cite{Ratra:1991bn,Carroll:1991zs,Garretson:1992vt,Field:1998hi,Lee:2001hj,Campanelli:2005ye,Anber:2006xt}, for inflation and
(p)reheating after
inflation~\cite{Anber:2009ua,Barnaby:2010vf,Barnaby:2011vw,Barnaby:2011qe,Barnaby:2012tk,Adshead:2013qp,Adshead:2015pva,Notari:2016npn}, for moduli decay~\cite{Giblin:2017wlo}, and for the
relaxion~\cite{Hook:2016mqo}. Our mechanism differs from previous cases because we focus on smaller couplings of the
axion to the gauge fields, which requires multiple oscillations of the axion until the gauge field grow sufficiently to
influence the axion dynamics. In this regime the relevant modes of the
gauge field go through two separate phases, a tachyonic growth period
and a subsequent parametric resonance period. Thus, our mechanism
shares features both with tachyonic preheating and with more standard
preheating scenarios based on parametric resonant decays.

Even though the coupling responsible for this mechanism is quite generic for axion-like particles, 
the presence of light charged
degrees of freedom can make the production of gauge fields very
inefficient due to plasma screening effects. For this reason we focus
on the coupling of the axion to a dark photon instead of making use
of its couplings to the SM photon.


We show that this mechanism can be very efficient in transferring
energy from the axion into the dark photons. As long as
backscattering effects of the dark photons into
the axions are small, 
this mechanism can open up most of the large-$f_a$
parameter space. We will briefly comment on potential limitations
of this mechanism due to the possibility  
of backscattering effects introducing significant deviations in
our results. A quantitative understanding of these effects
requires a dedicated numerical investigation beyond the scope of this
work.

There are other ways to evade the upper bound on $f_a$. One
possibility is that the initial misalignment angle 
is tuned to be small, with the tuning justified by anthropic
selection of the dark matter density~\cite{Linde:1987bx,Tegmark:2005dy}.
Another possibility is that the universe experienced entropy
production between the QCD phase transition and Big Bang
Nucleosynthesis~\cite{Dine:1982ah,Steinhardt:1983ia,Lazarides:1990xp,Kawasaki:1995vt}.
This mechanism can relax the bound to $f_a \lesssim 10^{15}\GeV$. 
If the axion potential was temporarily much larger in the very early
universe, such that $m_a > H$, it would relax to its
minimum earlier~\cite{Dvali:1995ce,Choi:1996fs,Banks:1996ea,Banks:2002sd}.
However, a viable model that generates a large axion potential is typically
quite involved and needs to explain why the early time minimum is
aligned with the minimum today.
There are other possible signatures of our mechanism, which can be
potentially used to distinguish it from these other mechanisms.
\begin{itemize}
  \item 
For
very large $f_a$, the energy in the axion is so large that, once it is
transferred to radiation, it can lead to observable changes in the
number of extra relativistic degrees of freedoms ($N_\text{eff}$)
during BBN and during the CMB decoupling. Current constraints on
$N_\text{eff}$ limit $f_a \lesssim 2 \times 10^{17}$ GeV for an axion with
order one initial misalignment angle. Future measurements are expected
to improve the bound on $N_\text{eff}$ by an order of magnitude which
will provide indirect tests of this mechanism in a large range of the
interesting parameter space. 
\item 
A dynamical mechanism that decreases the axion abundance also
significantly weakens the bounds on large-$f_a$ axions from the
non-observation of isocurvature perturbations in the CMB. This opens
up a sizable fraction of the parameter space for inflationary models,
specially for higher scales of inflation. If future experiments
measure a large scale for inflation and also discover axion DM with a
large $f_a$, this would be a strong indirect evidence against an
anthropic axion scenario. 
\item
The particle production mechanism might
also lead to significant changes in the small scale matter power
spectrum. This is due to a combination of effects which we describe in
the text. One of the main effects is that it
decreases the average energy density in the axion field without a
corresponding decrease in the energy density of its fluctuations at
scales that enter the horizon before particle production
becomes relevant. This amplifies the density perturbations generated by
inflation and enhances the power
spectrum at these small scales.
\end{itemize}

Even though the most economical possibility is that a single axion
field solves the strong-CP problem as well as constitutes all of dark
matter, there may be a plenitude of light
axions~\cite{Arvanitaki:2009fg}, since axions arise readily in string
theory compactifications~\cite{Banks:2003sx,Svrcek:2006yi,Conlon:2006tq}.
While in this paper we will focus on the QCD axion, our results can be
easily adapted to other axion-like particles.

After a brief review of axion dark matter in \cref{sec:axionreview}, we
will present a simple model of axions coupled to the dark photon with
appropriate strength to produce our effect in
\cref{sec:models}. We then describe the mechanism of particle
production and our results in \cref{sec:particleprod}, and we lay out
the future directions and conclusions in \cref{sec:conclusions}.

\section{Axion dark matter review}
\label{sec:axionreview}

The cosmology of the axion depends sensitively on whether the spontaneous
breaking of PQ symmetry occurs before/during or after inflation. In
this work we are interested in the case where the PQ symmetry is
broken before inflation, which is the case when the scale of PQ
breaking is larger than both Hubble during inflation and the reheat
temperature. Due to the
astrophysical constraints requiring $f_a \gtrsim 10^9$ GeV, and the
bound on the inflationary scale from non-detection of gravitational
B-modes in the CMB~\cite{Ade:2015lrj,Kleban:2015daa}, $H_I \lesssim
10^{14}$ GeV, this scenario represents a large fraction of the allowed
axion parameter space. In this case, due to the exponential expansion
during inflation, the axion field takes on a constant value throughout
our Hubble volume, which is generically away from the late time
minimum of the potential. The dark matter
abundance then arises out of this misalignment, and depends on the axion
mass and on the size of the misalignment.

\subsection{The axion Lagrangian and potential}
We review the axion couplings to the SM.
Following \cite{diCortona:2015ldu} the relevant terms in the axion
Lagrangian are given by 
\begin{align} 
  \mathcal{L}
  = 
  \frac12 (\partial_\mu \phi)^2
  +\frac{\alpha_s\phi}{8 \pi f_a} 
   G^a_{\mu \nu} \tilde{G}^{a\,\mu\nu} 
   + g_{a\gamma\gamma} \phi {F}^{\mu \nu}  {\tilde{F}}_{\mu\nu}
  +  \frac{\partial_\mu  \phi } { 2 f_a }   \sum_{\psi} c_\psi  (\bar{\psi}\gamma^\mu \gamma^5 \psi)
  \label{eq:axionlag}
\end{align} 
where $\tilde{G}^{\alpha\beta} = \frac12
G_{\mu\nu}\epsilon^{\mu\nu\alpha\beta}$, and $\psi$ are SM fermions.
We use the coupling to gluons 
as the definition of $f_a$, and the couplings $c_i$ to other SM fields 
are model-dependent. The coupling of the axion to photons is relevant
for many experimental searches, and for the KSVZ
axion~\cite{Kim:1979if,Shifman:1979if} is given by
\begin{align}
  g_{a\gamma\gamma}= \frac{-1.92 \alpha}{8\pi f_a}
  \,.
  \label{eq:ksvz}
\end{align}

The coupling of the axion to QCD breaks the continuous shift symmetry of the
axion to a discrete shift symmetry. 
For the axion to solve the
strong-CP problem, it should get its potential dominantly from QCD
dynamics.
The zero temperature potential is a low energy observable and can be
calculated directly from the chiral Lagrangian (see
e.g.~\cite{diCortona:2015ldu}),
\begin{align}
  V(\phi)
  &=
  -m_\pi^2 f_\pi^2 
  \sqrt{1 - \frac{4 m_u m_d}{(m_u+m_d)^2} \sin^2 \frac{\phi}{2f_a}}
  \approx 
  -m_\pi^2 f_\pi^2 
  +\frac{m_\pi^2 f_\pi^2}{2 f_a^2} 
  \frac{m_u m_d}{(m_u + m_d)^2} \phi^2 
  + \mathcal{O}(\phi^4/f_a^4)
\end{align} 
for small $\phi$ displacements.

At high temperature, the axion potential is suppressed. There are many
computations of the high temperature axion mass (dilute instanton gas,
lattice, instanton liquid model), which do not all
agree with each other~(see
e.g.~\cite{Berkowitz:2015aua,Borsanyi:2015cka,Bonati:2015vqz,Frison:2016vuc,Borsanyi:2016ksw,Dine:2017swf} for recent results and discussion). This is an important
open question which has
implications for the axion dark matter abundance, but which does not
change the conclusion that large $f_a$ axions generically lead to an
overabundance of dark matter. For clarity, we
parametrize the mass in the following simple way which resembles the
result from the dilute instanton gas
calculation~\cite{Fox:2004kb,Hertzberg:2008wr},
\begin{align}
  m_a(T) 
  &= 
  \left\{
    \begin{array}{lr}
      m_a(0) \ , & T < 200\MeV\\
      b \, m_a(0)  \left(\frac{200\MeV}{T}\right)^4 \ , & T \geq 200\MeV\\
    \end{array}
    \right.
\label{eq:maT}
\end{align}
where $m_a(0) = (78\MeV)^2/f_a$ and $b = 0.018$. Note that in our approximation the mass is discontinuous, which is not expected to be actual behavior for the axion mass. The simple power law dependence in the temperature is expected to hold at large temperatures, $T \gtrsim 1 \GeV$, where the dilute instanton gas approximation is reliable. Given the disagreement in how to estimate the axion potential closer to the QCD phase transition scale, we extended the higher temperature behavior past the regime where it is reliable in order to have a concrete formula. As already discussed this introduces uncertainties in the calculation for the axion abundance but does not change the qualitative picture which is what we want to focus on.

\subsection{Axion cosmology}

Axions are produced in the early universe via thermal production
through its couplings to the SM plasma \cite{Masso:2002np,Graf:2010tv,Salvio:2013iaa} 
and non-thermally via the
misalignment mechanism \cite{Preskill:1982cy}. Thermally produced axions have momenta
comparable to the temperature of the SM plasma and thus are an extra
form of radiation. The abundance of thermal axions is negligible for
large $f_a$, which is the case we are interested in. The non-thermal
production generates a coherent oscillating field, which behaves as a
non-relativistic fluid and can be the dominant contribution to the
dark matter density. The relic density is determined by the initial
misalignment angle of the axion, $\theta_i = \phi_i/f_a$. 
The axion starts to oscillate when $m_a
\sim H$, and can be subsequently described as a bath
of zero momentum particles. The axion abundance can be easily
calculated by using the fact that shortly after oscillations begin the
axion co-moving number density is conserved. 
If $f_a > 2 \times 10^{17}\GeV$,
the axion starts to oscillate at $T_{osc} <  \Lambda_{QCD} \simeq
200\MeV$, when the
axion mass is roughly constant.
For  $f_a < 2 \times 10^{15}\GeV$ the oscillations begin at a
higher temperature, $T_{osc} \gtrsim 1 \GeV$, where the axion mass depends
sensitively on the temperature and is well approximated by \cref{eq:maT}.
In these two regions, the axion DM relic density is
\cite{Fox:2004kb},
\begin{align} 
  \Omega_{a} h^2 
  \sim \ 
  \left\{ \begin{array}{ll} 
    2 \times 10^{4}  
    \left(
    \frac{f_a} {10^{16}\GeV} 
    \right)^{7/6} \theta_i^2 
    \, , &  f_a < 2 \times 10^{15} \GeV     
    \\
    5 \times 10^{3}  
    \left(\frac{f_a} {10^{16}\GeV} \right)^{3/2}  \theta_i^2 
    \, , &  f_a > 2 \times 10^{17}\GeV 
  \end{array} \right.
  \label{eq:axion-abundance}
  \,,
\end{align} 
where we have kept only the quadratic piece of the axion potential. 
In the region $2\times10^{15} \GeV < f_a < 2\times10^{17} \GeV$, the
axion begins oscillation around the QCD scale,
$T_{osc} \sim \Lambda_{QCD}$. In this region neither expression above is
accurate and strong QCD effects affect the axion
abundance. For our numerical results we calculate the axion abundance
explicitly by numerically evolving the axion condensate using the
temperature-dependent mass in \cref{eq:maT}.
Using the estimate above, we see that for $f_a \sim 10^{15}\GeV$, we
would need an initial misalignment $\theta_i^2 \sim 10^{-4}$
to get the observed relic density.

Another important constraint on axion dark matter comes from the limit
on isocurvature perturbations. Because the axion is effectively a free
scalar field during inflation, it experiences scale invariant
fluctuations due to the de Sitter background  $\delta \phi  \sim
H_I / (2 \pi ) $. These correspond to isocurvature perturbations and
are constrained from CMB measurements. The isocurvature power
spectrum due to axion fluctuations is given by 
\begin{equation} 
   P_{iso} =  
   \frac{\Omega_a^2}{\Omega_\text{DM}^2}
   \frac{ H_I^2} { \pi^2 \, f_a^2 \, \theta_i^2 } \ , 
   \label{eq:isocurvature-power}
\end{equation} 
where we allowed the axion abundance to differ from the observed dark matter abundance $\Omega_\text{DM}$. 
This value depends on the $\theta_i$, because the perturbations are
defined by $\delta \rho_a/\rho_a$, and the average density of the
axions depends on the misalignment angle.

\begin{figure}[t]
\begin{center}
\includegraphics[scale=0.6]{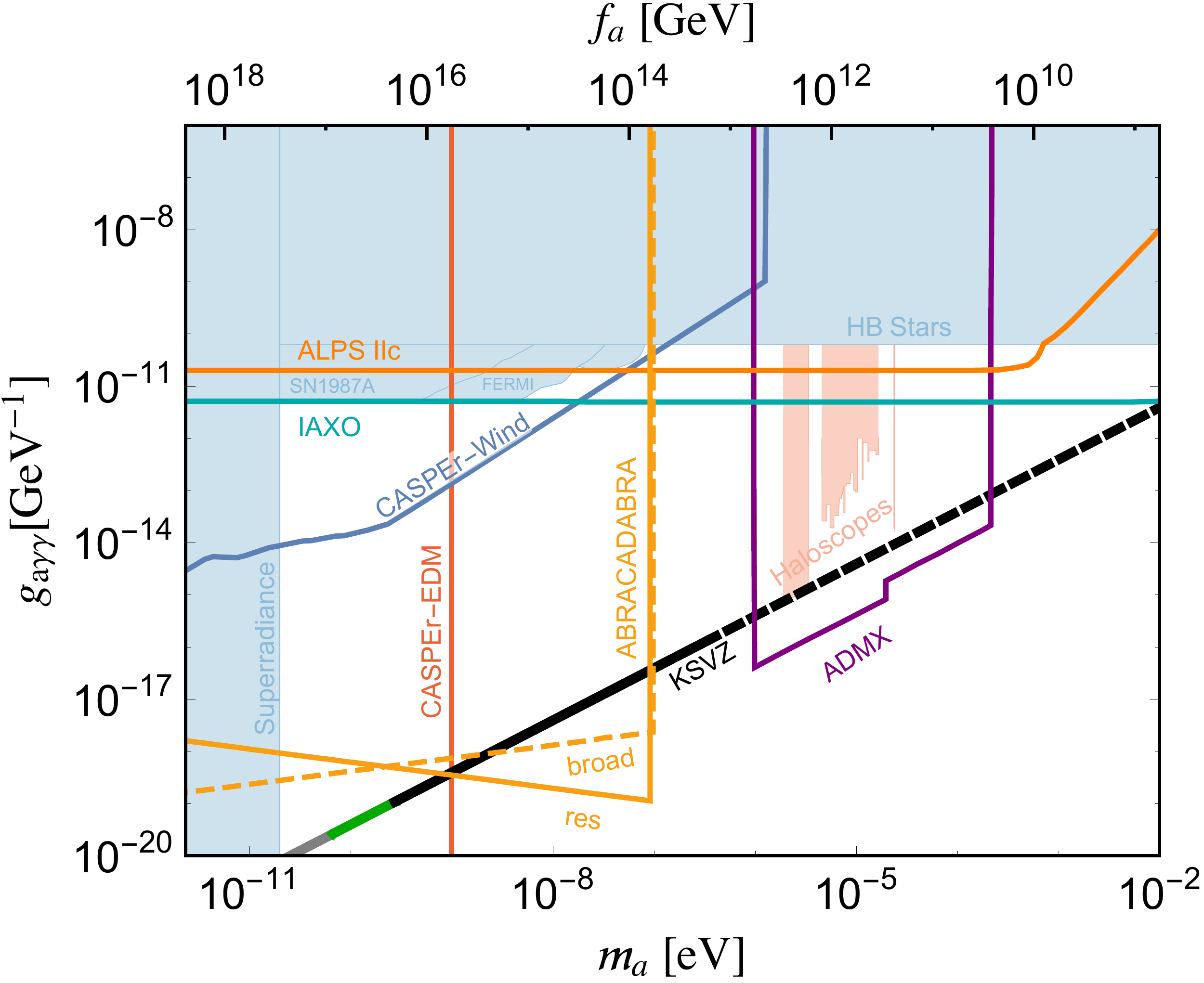}
\caption{Current bounds and future experimental reach for axion
  couplings to photons defined in
  \cref{eq:axionlag}.
  Blue shaded regions show the current constraints which are independent of a dark matter
  interpretation, while the red shaded region are constraints from
  axion haloscopes. Sensitivity of future experiments is shown by
  colored lines.
  The thick black line shows the prediction for the KSVZ axion as
  in~\cref{eq:ksvz}.
  The solid part of the line indicates where the initial misalignment
  would need to be tuned $\theta_i < 0.1$ or worse, which is
  where our mechanism is needed to have the correct axion abundance.
  For larger values of $f_a$ (the green part of the line) the extra
  dark radiation produced
  by our mechanism  will be within the
  sensitivity of future CMB experiments. Even larger $f_a$ values (the
  gray part), are already constrained by Planck observations (see
  discussion in~\cref{sec:neff}).
}
\label{fig:axion-reach}
\end{center}
\end{figure}

The CMB has measured the scalar perturbations as predominantly adiabatic perturbations
with a scalar power spectrum, $P_s \simeq 2 \times 10^{-9}$, where
we neglect the spectral index. With the assumption that the scalar
power spectrum is uncorrelated with the axion isocurvature 
perturbations, CMB constrains~\cite{Ade:2015lrj}, 
\begin{align}
r_{iso} = P_{iso} / P_s < 0.038 \, .
\end{align}

The bound on isocurvature perturbations leads to an upper bound on Hubble during inflation
\begin{align}
  \frac{H_I}{f_a} 
  &\lesssim 10^{-5} \theta_i \frac{\Omega_\text{DM}}{\Omega_a} \, ,
\end{align}
If axion constitutes all of dark matter from \cref{eq:axion-abundance}
one sees that for large $f_a$ one requires $\theta_i \ll 1$, which
makes the isocurvature bound on the inflationary scale stronger.
Therefore if the initial misalignment angle is $\mathcal{O}(1)$, and a
dynamical mechanism decreases the axion abundance to the observed DM
abundance, the isocurvature bound is weakened. This is the case for
the late entropy production
scenarios~\cite{Visinelli:2009kt,Co:2016vsi}, and is also the case for
the particle production mechanism we explore in this work. For other recent ideas that 
relax the axion isocurvature bounds see e.g.~\cite{Kawasaki:2015lpf,Nomura:2015xil,Kearney:2016vqw}.

There are a number of tabletop experiments and astrophysical
observations that put constraints on the viable axion parameter space.
There are also new experimental proposals which 
have the potential to reach the QCD axion in a wide mass window in the
near future. 
The existing limits and future sensitivity on the axion coupling to photons are summarized in
\cref{fig:axion-reach}. We highlight the light axion region, which is
the region of interest for new experiments. This region would be ruled
out for generic initial conditions, but our mechanism makes the entire
region viable.
The blue and red shaded regions in this figure represents the
current bounds. 
It includes the cavity microwave experiment,
ADMX~\cite{Asztalos:2010}, which has probed axion DM of about
$\SI{}{\micro\eV}$; the helioscope, CAST~\cite{Andriamonje:2007ew}, to
search for solar axion coupling to photons; the supernova SN1987A
bound~\cite{Payez:2014xsa} from the Primakoff process;
FermiLAT~\cite{TheFermi-LAT:2016zue}, H.E.S.S
~\cite{Abramowski:2013oea} and X-ray 
observations~\cite{Berg:2016ese,Marsh:2017yvc,Conlon:2017qcw} bounds from the
irregularity of the photon spectrum;
globular clusters~\cite{Ayala:2014pea}; black hole
superradiance~\cite{Arvanitaki:2014wva} (ignoring the disruption of the
axion cloud due to interactions at larger $g_{a \gamma \gamma}$).  We show the
projected sensitivity of future experiments,
ABRACADABRA~\cite{Kahn:2016aff},
ALPS II~\cite{Bahre:2013ywa},
ADMX~\cite{Shokair:2014rna}, CASPEr-Wind and
CASPEr-Electric~\cite{Budker:2013hfa}, and the helioscope
IAXO~\cite{Armengaud:2014gea}.

\section{A simple model for particle production}
\label{sec:models}
In addition to the standard QCD axion Lagrangian discussed above, we
assume an additional coupling of the axion to a dark photon (such
couplings have recently been explored in a different
context~\cite{Kaneta:2016wvf,Kaneta:2017wfh}),
\begin{align}
  \mathcal{L}
  &=
 \frac{\alpha_d c_d}{8\pi f_a} \phi\,F_D^{\mu\nu} (\widetilde{F}_D)_{\mu\nu}
\end{align}
As we will see below,
the particle production mechanism is only effective when
the coupling of the axion to the dark photon is 
larger than the coupling to QCD,
with $r \equiv \frac{c_d\alpha_d}{2\pi}\sim \mathcal{O}(10)$.
This hierarchy of couplings is easily seen to be stable under quantum
corrections at the effective theory level. Nevertheless, one may
wonder how easy it is to obtain this structure from a UV model. 
There are a number of possible model building avenues. We
will not pursue them in detail here, but present a simple example as a
proof of principle.

One way to generate a hierarchy in axion couplings is using the
Kim-Nilles-Peloso (KNP) alignment mechanism~\cite{Kim:2004rp,Dvali:2007hz,Choi:2014rja,Bai:2014coa,delaFuente:2014aca,Choi:2015fiu,Kaplan:2015fuy,Giudice:2016yja,Farina:2016tgd,
Craig:2017cda}. 
As a specific case, we take two axions, say $a$ and $b$, each coupling
separately
to the two gauge groups, QCD and the dark photon, respectively.
We include an additional ``aligned''
contribution to the axion mass matrix which breaks the two
shift-symmetries down to a single shift-symmetry such that one linear
combination of $a$ and $b$ gets a large mass and decouples. If the
light eigenstate has a hierarchical overlap with $a$ vs.~$b$, then it
couples to QCD and the dark photon with a hierarchical coupling.

The effective Lagrangian is,
\begin{align}
  \mathcal{L}
  &=
  \frac{\alpha_s}{8\pi f} a \, G^{a,\mu\nu} \widetilde{G}^a_{\mu\nu}
  +
  \frac{\alpha_d}{8\pi f} b \,F_D^{\mu\nu} (\widetilde{F}_D)_{\mu\nu}
  +\Lambda^4 \cos \left(\frac{n a + b}{f}\right)
\end{align}
where we have included a simple example of the explicit symmetry
breaking term that we want.
This symmetry breaking term can be UV-completed in a number of different
ways, which we discuss in \cref{app:uvmodel}.

The eigenstates from the above Lagrangian are,
\begin{align}
  \phi = \frac{1}{\sqrt{1+n^2}} (a-n b),
  \qquad 
  \phi_h = \frac{1}{\sqrt{1+n^2}} (n a + b)
\end{align}
where $\phi$ is the QCD axion, and $\phi_h$ is the heavy mass
eigenstate.
We can thus rewrite the effective Lagrangian in terms of the light eigenstate,
\begin{align}
  \mathcal{L}
  &=
  \frac{\alpha_s}{8\pi f_a} \phi \, G^{a,\mu\nu} \widetilde{G}^a_{\mu\nu}
  +
  \frac{\alpha_d n }{8\pi f_a} \phi  \,F_D^{\mu\nu} (\widetilde{F}_D)_{\mu\nu}
  \label{eq:axionefflagrangian}
\end{align}
with $f_a = f\sqrt{1+n^2}$. In other words, $c_d = n$, or 
$r = n \alpha_d / (2\pi)$.

\section{Particle production and $\Omega_{DM}$}
\label{sec:particleprod}

We now study the changes in the axion evolution due to its coupling to
the dark photon. In our setup, the QCD axion
couples to dark photons, with a coupling that is $\sim
\mathcal{O}(10)$
times larger than its coupling to the standard model particles. 
This coupling is responsible for depleting the axion abundance.

If the mass of the dark photon exceeds the mass of the axion,
particle production from axion decays become ineffective.
This puts a severe constraint
on the mass of the dark photon; the simplest possibility is that the
dark photon is massless.
We also assume that the dark photon is in its vacuum state, with no
light matter charged under it in the universe,
when the axion
oscillations start around the QCD phase
transition.
If there is a non-negligible abundance of matter charged under the
dark photon it generates a Debye mass for the dark photon,
shutting off particle production.
The presence of this Debye mass is
precisely why the axion coupling to the photon cannot be utilized for
our mechanism. The existence of light degrees of freedom charged under the dark
photon would also lead to large self-interaction for the dark photons which would also
suppress the particle production effect. Thus, we assume that matter
charged under the dark photon is heavy, which is the natural
expectation if it is a scalar or a vectorlike fermion.

\subsection{The equations of motion}
The equation of motion (EOM) for the homogeneous axion field in the
Friedmann-Robertson-Walker cosmological 
background with the scale factor $a(\eta)$ is written as 
\begin{align}
  \phi'' + 2 a H \phi'
  + a^2 \frac{\partial V}{\partial \phi}
  =
  \frac{a^2 }{4 f_d} (F_D)_{\mu \nu} \tilde{F}_D^{\mu\nu} \, ,
  \label{eq:eom-axion}
\end{align}
where the prime denotes derivative with respect to conformal time $\eta$ and we defined $1/ f_d =  \alpha_d c_d / 2 \pi f_a$. Working in Coulomb gauge, $\vec \nabla \cdot \vec A = 0$, the equations of motion for the gauge fields are
\begin{align}
A_0 & = 0 \\
\vec A'' 
& - \nabla^2 \vec A 
= -\frac{\phi'}{f_d} \, \vec \nabla \times \vec A
\label{eq:gauge-eom}
\end{align}

We treat the homogenous axion field as a classical field and focus on
the production of gauge field quanta due to the coupling between the
axion and gauge field. To see this particle production effect we
decompose the gauge field in terms of creation and annihilation
operators
\begin{align}
  A(\eta,\vec{x})
  &=
  \sum_{\lambda=\pm}
  \int\frac{d^3 k}{(2\pi)^3}
  a_\lambda^{\vec{k}} A_\lambda(\eta,\vec{k}) \epsilon_\lambda e^{i
    \vec{k}\cdot\vec{x}}
  +
  \rm{h.c.}
\end{align}
where $\lambda = \pm$ are the helicities of the gauge field. Plugging this decomposition into \cref{eq:gauge-eom} we find
\begin{align}
  A_\pm'' + \left(k^2 \pm \frac{k \phi'}{f_d}\right) A_\pm &=0
\end{align}

We fix the initial conditions for the gauge fields to be consistent with the Bunch-Davies vacuum
\begin{align}
A_\pm(\eta \ll \eta_{\text{QCD}}) \rightarrow \frac{e^{- i k \eta}}{\sqrt{2 k}}
\end{align}
with $\eta_\text{QCD}$ the conformal time of the QCD phase transition. The energy density in the gauge field is given by
\begin{align}
\rho_d
  = \frac{1}{2a^4} 
  \int \frac{d^3k}{(2 \pi)^3} 
  \left(
  \left|\frac{\partial A_+}{\partial \eta} \right|^2
  +\left|\frac{\partial A_-}{\partial \eta} \right|^2
  +k^2( |A_+|^2 + |A_-|^2)
  -2k
  \right)
 \label{eq:gauge-energy}
\end{align}
where the last term comes from subtracting the vacuum energy, to
render the integral finite. From this expression we see that if some
modes of the gauge field described by \cref{eq:gauge-eom} experience a
large growth this leads to an increase in the energy of the radiation
field. This energy gain corresponds to particles being generated by
the coupling to the time dependent classical axion field.

By energy conservation, the increase in energy of the radiation field
leads to a decrease in energy of the axion. To study the backreaction
effects of the particle production in the dynamics of the axion we
will treat the right-hand side of \cref{eq:eom-axion} as a mean field
\begin{align}
(F_D)_{\mu \nu} \tilde{F}_D^{\mu\nu}
  \rightarrow 
  \langle 
  (F_D)_{\mu \nu} \tilde{F}_D^{\mu\nu}
  \rangle
  &= 
  \frac{1}{a^4} 
  \int \frac{d^3k}{(2 \pi)^3} 
  \frac{|\vec k|}{2} 
  \frac{\partial}{\partial \eta} 
  \left( |A_+|^2 - |A_-|^2 \right) \, .
  \label{eq:edotb}
\end{align}

\subsection{Conditions for efficient particle production}
These coupled differential equations can be solved numerically to
study the evolution of the axion. Before presenting our numerical
results, we first discuss when we would expect particle production to
be efficient. For this qualitative discussion we ignore the temperature
dependence of the axion mass and the change in the number of
relativistic degrees of freedom during its evolution. We include these
effects for our numerical analysis in the next section.

For simplicity we approximate the axion potential by keeping only its
leading quadratic term 
\begin{align}
V(\phi) \approx \frac{1}{2} m_a^2 \phi^2 + \dots
\end{align}
which is a good approximation for $\phi \lesssim f_a$. 
The axion starts oscillating once Hubble drops below the axion mass
\begin{align}
  \phi 
  \approx 
  \phi_i \cos\left( m_a t  \right) \, 
  \left( \frac{a_i}{a} \right)^{3/2}
\end{align}
where $t$ is the comoving time and $a_i$ is the scale factor when $H =
m_a$. This equation holds as long as the backreaction due to the
gauge fields is small.

The fact the coupling in \cref{eq:axionefflagrangian} can lead to a
very large production of the gauge field can be seen directly from the
equations of motion for the gauge field
\begin{align}
  A_\pm'' + \omega^2(k, \phi') A_\pm = 0 
  ,\qquad
  \omega^2(k, \phi') 
  = 
  k^2 \pm \frac{k \phi'}{f_d}
  \,.
\label{eq:tachyon-frequency}
\end{align}
If the effective frequency $\omega^2$ of a given mode becomes
negative, the corresponding mode becomes tachyonic which can lead to
an exponential growth of the corresponding ``mode function''.
From \cref{eq:tachyon-frequency} one sees that the largest tachyonic
frequency is
\begin{align}
  |\omega_t|
  \lesssim
  \frac{\phi'}{2 f_d}
  \approx 
  \frac{a m_a \phi_i}{2 f_d}
  \left( \frac{a_i}{a} \right)^{3/2} 
  \leq
  a m_a \frac{r \theta_i}{2} 
\end{align}
where we have used the dimensionless variable $\theta =
\phi/f_a$, and recall that $r = f_a/f_d$.
The largest tachyonic frequency occurs near the start of the
oscillations, before the amplitude of $\phi$ decreases due to Hubble
friction. 

In
order for the exponential growth to occur, another condition must be
satisfied: the corresponding growth rate, $\sim |\omega|$, needs to be
fast compared to both the Hubble\footnote{ Note that since we are
  using conformal time, one should compare the tachyon rate to $a H$
instead of just $H$.  } expansion rate and to the time scale for
which the effective frequency $\omega^2$ remains negative. When both
conditions are satisfied, the growth of the mode $A(k)$ can be
estimated using a WKB approximation
\begin{align}
  A \propto \exp\left( \int d\tilde{\eta}\, |\omega| \right)
\end{align}
where the $\tilde{\eta}$ integration is only over times in which the mode is
tachyonic.

A given frequency mode remains tachyonic over a half-period of the
axion, before $\phi'$ changes sign. Thus, the
conformal time
interval in which a mode is tachyonic is approximately given by
$\Delta \eta_\text{tachyon} \simeq 1/(a m_a)$.
There is no real growth of the gauge fields when 
$|\omega| \ll (\Delta \eta_\text{tachyon})^{-1}$.
In this case, the time scale associated with the axion oscillation
is too fast compared to the typical time scale for tachyon growth, and
therefore the gauge field modes effectively respond to an averaged
axion field, and averaging $\phi'$ over an oscillation is almost
zero. 
Thus, the lower bounds on the frequency that experiences an exponential
production is
\begin{align}
  \begin{aligned}
    |\omega_t| 
    \gtrsim & a H 
    \\
    |\omega_t| 
    \gtrsim & (\Delta \eta_\text{tachyon})^{-1} \approx a m_a 
  \end{aligned}
\end{align}
The second condition is stronger than the first since $H\leq m_a$. 
Therefore, the gauge modes are exponentially produced in the
following frequency band,
\begin{align}
  a m_a 
  \lesssim
  |\omega_t| 
  \lesssim
  a m_a \frac{r \theta_i}{2} 
  \label{eq:tachyon-conditions}
\end{align}
Given
that $f_a$ determines the range for the axion field, the initial
misalignment angle $\theta_i$ is bound to be at most $\mathcal{O}(1)$.
Therefore, in order to have a significant production of gauge fields,
one must have $f_d > f_a$. As we will
show numerically in the next section, the ratio $f_d/f_a$ controls how
effective the energy transfer between then axion condensate and the
gauge field is.

In the relevant parameter space the gauge fields do not grow enough
during the first oscillation of the axion field for them to influence
its dynamics. At each oscillation the range of $k$-modes that
experience tachyonic growth becomes smaller and eventually disappears.
Therefore tachyonic particle production is only relevant during the
first few oscillations of the axion. After a $k$-mode leaves the
tachyonic frequency band it can still experience a significant growth
sourced by the axion oscillations once its physical momenta redshifts
enough for its frequency to enter the parametric resonance band. The
growth due to the parametric resonant behavior is only relevant
because it is Bose-enhanced due to the initial exponential growth of
that mode during the tachyonic phase. The dynamics of the coupled
axion-gauge field dynamics is very rich and due to its intrinsically
non-linear nature it is very sensitive to variation of parameters and
initial conditions as will be shown in the results section. Once all
$k$-modes of the gauge field that experienced tachyonic growth redshift
to momenta smaller than the axion mass, the axion-gauge field coupling
becomes ineffective at depleting the axion energy, and the axion
subsequently evolves as in the standard case.


\subsection{Dark photon contribution to $N_\text{eff}$}
\label{sec:neff}
The dark photons produced due to the tachyonic instability will carry
a significant fraction of the energy density originally stored in the
axion field. Since the gauge bosons are massless their energy density
dilutes like radiation, and thus they contribute to the number of
relativistic degrees of freedom ($N_\text{eff}$) at BBN and also at
the time of CMB decoupling. Their contribution to $N_\text{eff}$ is
given by
\begin{align}
  \Delta N_\text{eff} 
  &= 
  \left.
  \frac{8}{7} \left( \frac{11}{4} \right)^{4/3} 
  \, \frac{\rho_d}{\rho_\gamma}
  \right|_{T_\gamma = 1\,\rm{eV}}
  \label{eq:neff}
\end{align}
where $\rho_\gamma$ is the energy density in photons and $\rho_d$ is
given by \cref{eq:gauge-energy}. The contribution to $\Delta
N_\text{eff}$ as a function of the parameters $f_a$ and $f_d$ can be
computed using the numerical solutions to the gauge field-axion
system, and will be presented in the next section.

Assuming that most of the energy originally in the axion field is
transferred to the dark photon we can obtain the approximate bound
from $\Delta N_\text{eff}$.
As one can see from the numerical results in the next section, the
transition between the gauge field feedback being negligible to it
being very important in the axion evolution is very fast. We also find
that this transition usually happens when the scale factor has
increased (temperature has decreased) by an $\mathcal{O}(10)$ factor
from when the oscillations started and that at this point an order one
fraction of the axion energy is transferred to dark photons.
Therefore, ignoring the temperature dependence of the mass (which is
a very good approximation for $f_a \gtrsim 10^{17}$ GeV), we find
\begin{align}
  \Delta N_\text{eff}  
  \sim 
  \frac{g_* \pi^2 \theta_0^2 f_a^2}{90 M_p^2} \, 
  \frac{a_\text{fb}}{a_\text{osc}} \, ,
\end{align}
where $g_*$ is the number of relativistic degrees of freedom in the SM
below the QCD phase transition and $a_\text{fb}$ is the scale factor
when the feedback due to the gauge fields becomes important for the
axion. Using this simple approximation we see that for $f_a \gtrsim
10^{17}$ GeV and $\theta_0 \sim 1$, the contribution to $N_\text{eff}$
becomes order one and is thus in tension with observations. In the
results section we present more precise bounds on $N_\text{eff}$
obtained from the numerical solutions.

\subsection{Numerical results}

\begin{figure}[thp]
  \centering
  \includegraphics[width=0.65\textwidth]{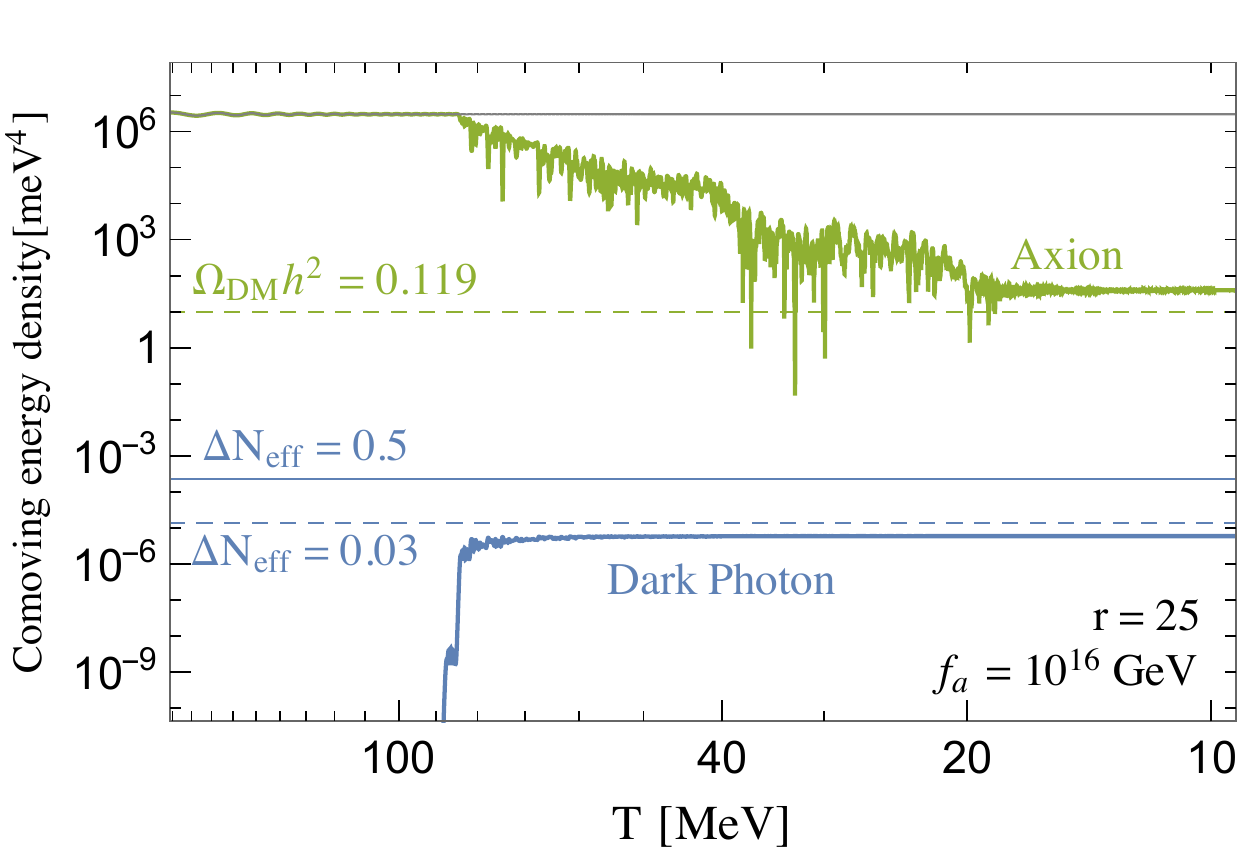}
  \\
  \vspace{10mm}
  \includegraphics[width=0.65\textwidth]{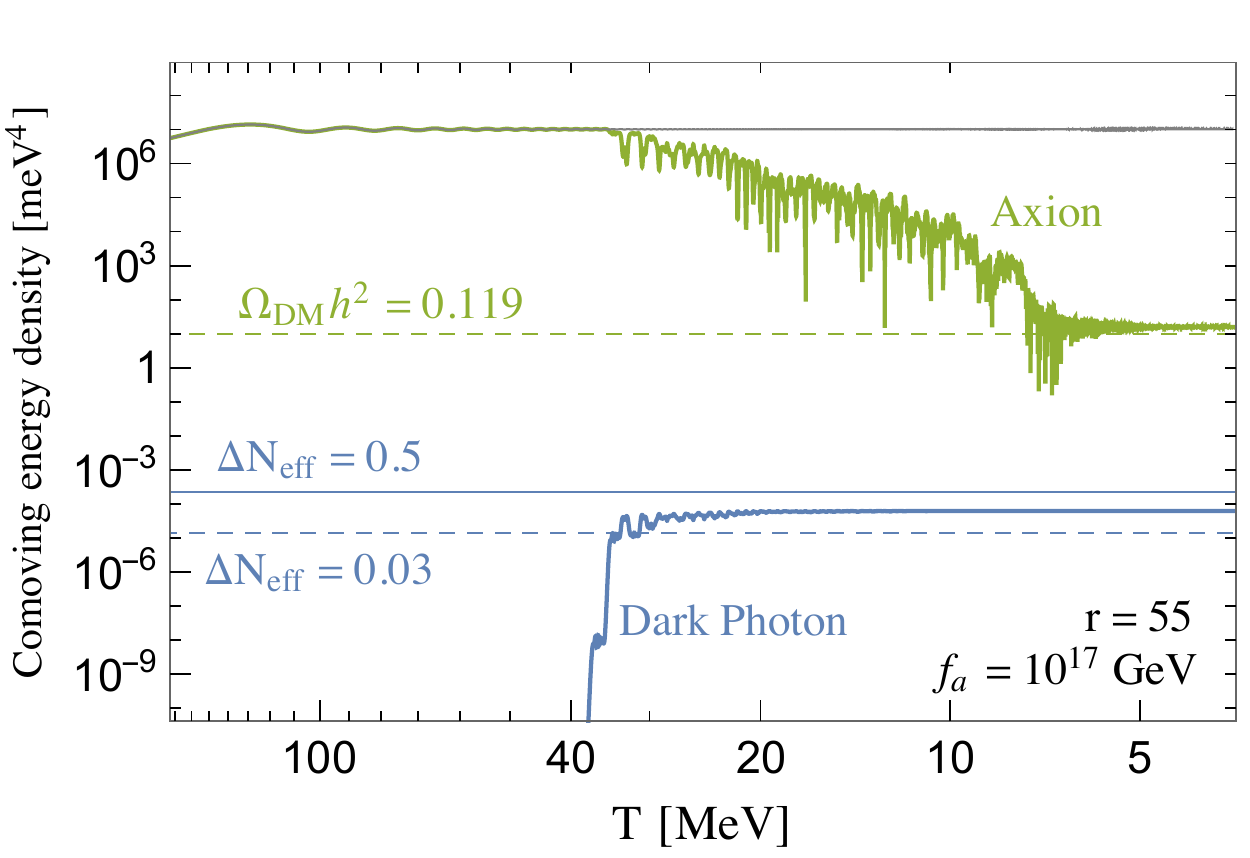}
  \caption{Comoving energy densities of the axion ($\rho_a a^3$, green) 
    and the dark
    photon ($\rho_d a^4$, blue) as a function of the temperature, with
    $a=1$ today,
    for two example values of $f_a$ and $r$. We also show the
    evolution of axion energy density in the absence of our mechanism
    (gray). The green dashed line indicates the correct dark matter
    abundance, and the blue horizontal lines indicate the current
    (solid) and
    future (dashed) sensitivity to the dark photon
    contribution to $\Delta N_{\rm{eff}}$. The absolute value of the
    axion and the dark
    photon energy densities are comparable at the point in which the axion energy density starts departing from the usual QCD axion behavior.
    The initial misalignment was chosen to be $\theta_i=1$.
  } 
\label{fig:depletion}
\end{figure}

\begin{figure}[t]
  \centering
  \includegraphics[width=0.65\textwidth]{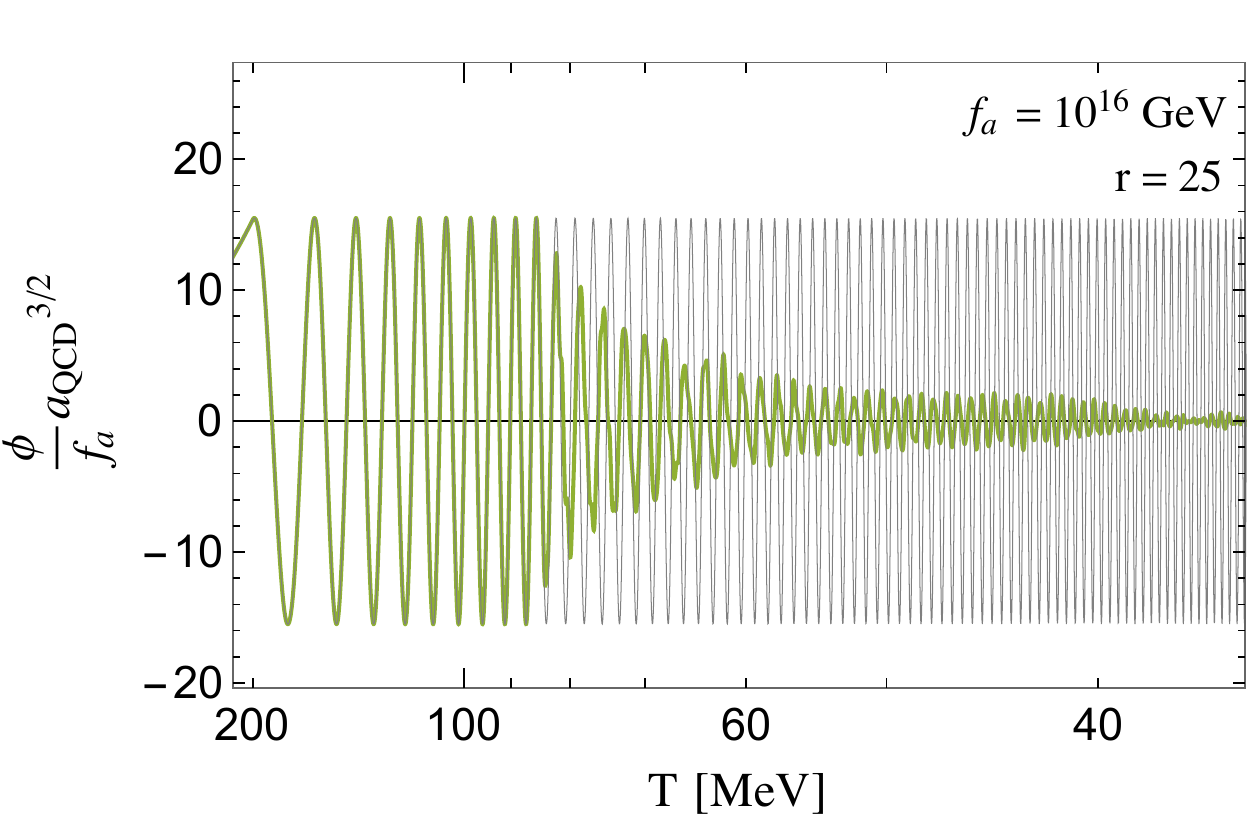}
  \\
  \vspace{10mm}
  \includegraphics[width=0.65\textwidth]{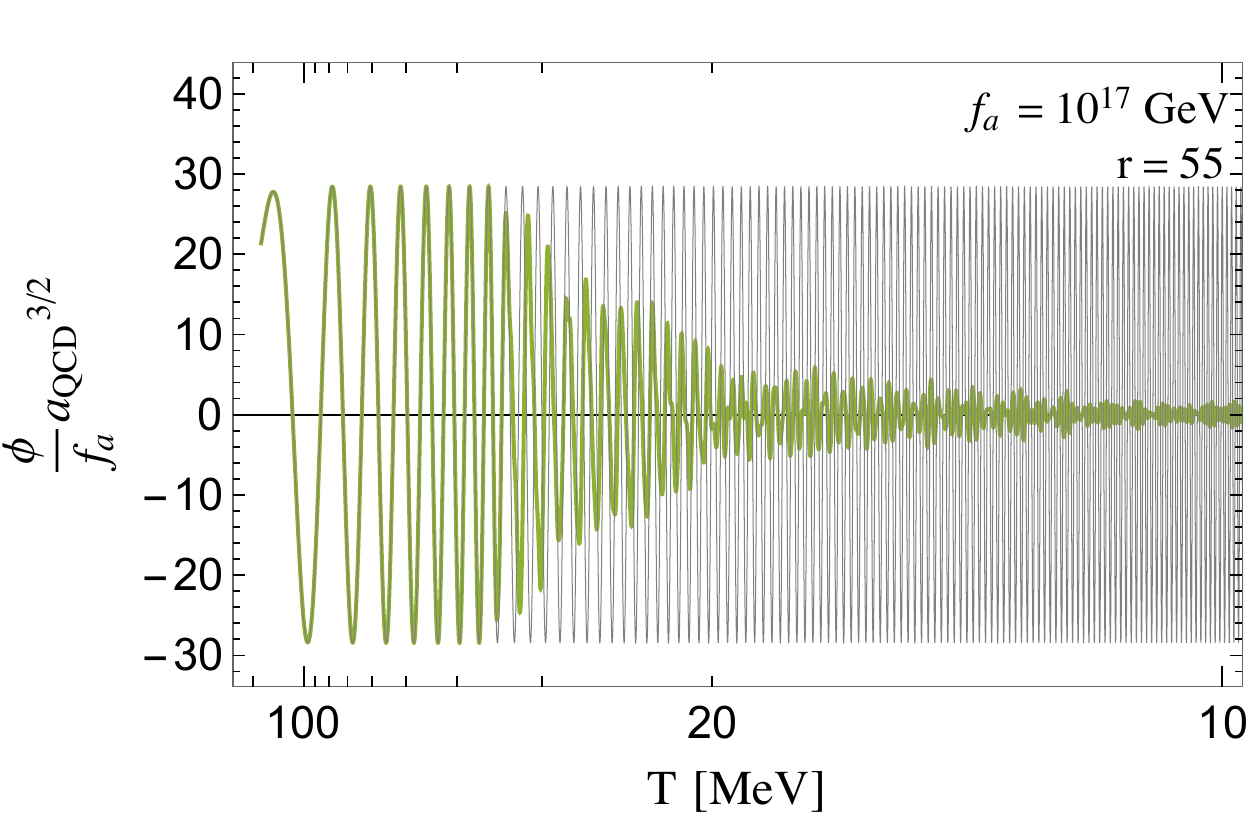}
  \caption{
    {The decay of the comoving axion amplitude (green) as a function of
    temperature due to particle production for two example values of
    $f_a$ and $r$, compared with the axion amplitude in the absence of
    our mechanism (gray). We normalize the scale factor $a_{QCD} (T) 
    \equiv \frac{a(T)}{a(1\GeV)}$.
The initial misalignment was chosen to be $\theta_i=1$.
} 
  }
\label{fig:phi}
\end{figure}

In this section we present our numerical results.
In order to produce
these results we
discretized the momentum integral in \cref{eq:edotb}. This was done by
introducing a maximum momenta $k_\text{cut} = 2 a_i m_a f_a/f_d$ and
discretizing the interval $(0, k_\text{cut})$ with over $10^4$ equally
spaced points, and solving the coupled axion-gauge-field system
until the particle production effect
becomes negligible. We tested the stability of the code
by varying the total number of discretized momenta, the maximum
momenta $k_\text{cut}$ and the precision of the algorithm.
For our numerical results we used the finite temperature potential of
\cref{eq:maT} and included the temperature dependence of the number of relativistic degrees of freedom~\cite{Olive:2016xmw}. This leads to analogous, but
quantitatively different
conditions from \cref{eq:tachyon-conditions}.

In \cref{fig:depletion,fig:phi} we show how particle
production affects the evolution of the axion energy density. After an
initial period where the axion abundance follows the standard axion,
the dark photons backreact on the axion, damping its energy density by
many orders of magnitude. We see that $f_a\simeq
10^{16}\GeV$ and $10^{17}\GeV$ are perfectly consistent
with measurement of $N_{\rm{eff}}$
by Planck~\cite{Ade:2015xua}. Larger decay constants would produce too
much dark radiation and hence are excluded.
Future CMB experiments~\cite{Abazajian:2016yjj} aim to measure $\Delta
N_{\rm{eff}}$ to $10^{-2}$ level, which will be sensitive to decay
constants of $f_a \gtrsim 10^{16}\GeV$. We
note that these results are for
an initial misalignment angle $\theta_i=1$, and the $N_{\rm{eff}}$
constraints are ameliorated for a smaller initial misalignment angle.

\begin{figure}[t]
  \centering
  \includegraphics[width=0.65\textwidth]{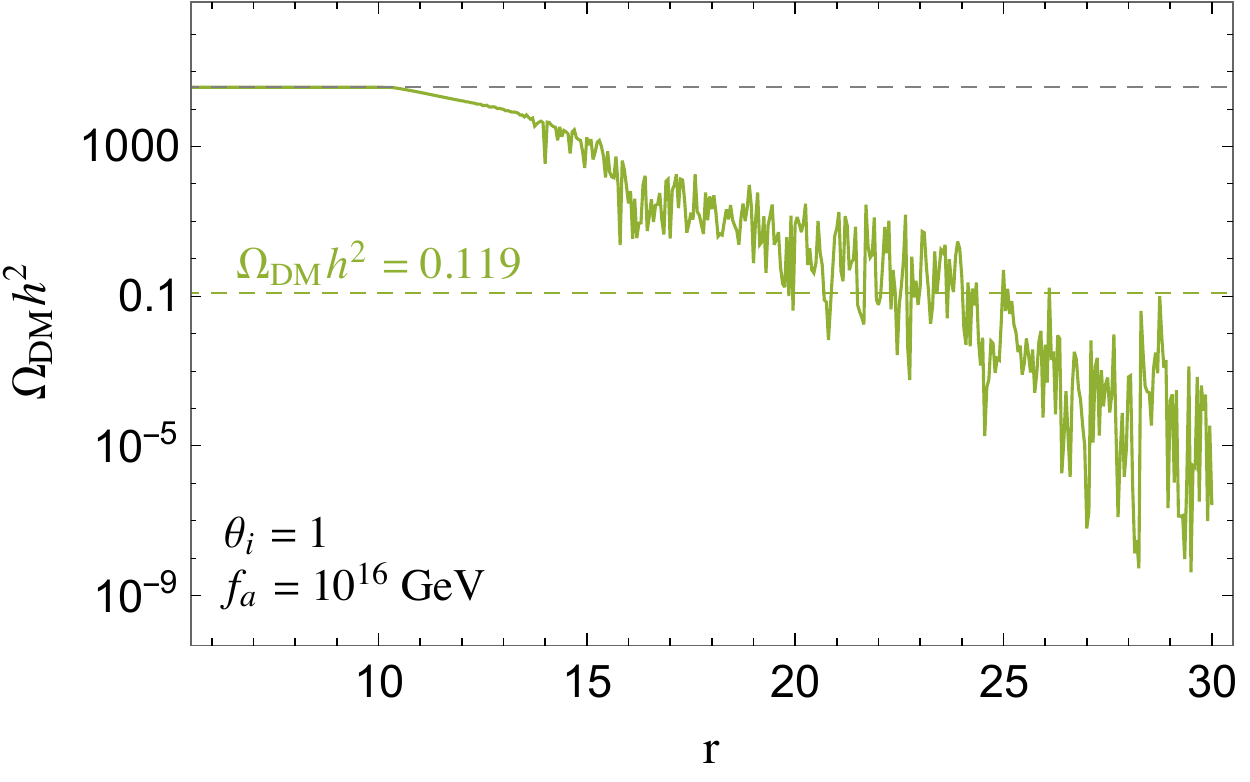}
  \\
  \vspace{10mm}
  \includegraphics[width=0.65\textwidth]{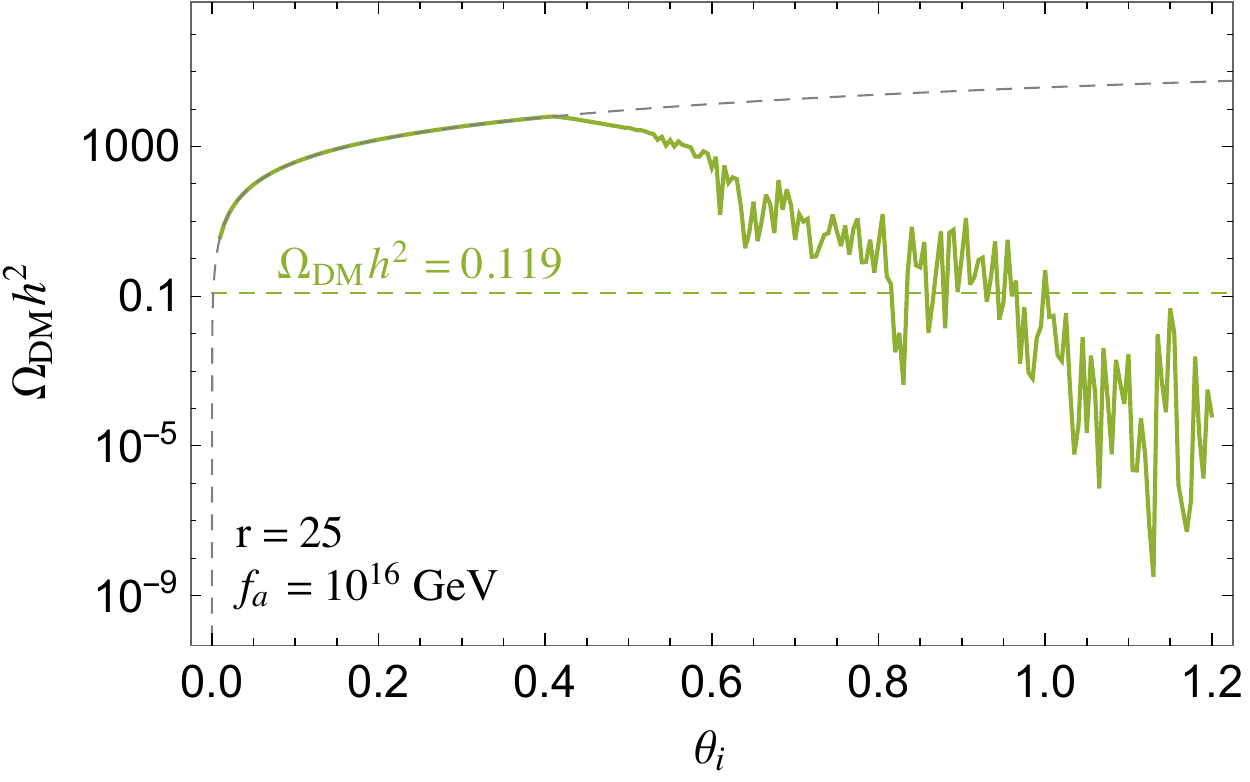}
  \caption{
    {The sensitivity of the axion abundance to the ratio $r = f_a/f_d$
    (left) and to the initial misalignment angle $\theta_i$ (right).
  The dashed gray line is the axion abundance in the absence of the
dark photon coupling. We see that the mechanism is ineffective for
small values of $r$ and $\theta_i$. For larger values of
$\{r,\theta_i\}$, the depletion is larger and very sensitive to the
initial conditions/parameters.} 
  }
\label{fig:sensitivity}
\end{figure}

In \cref{fig:sensitivity} we show that the depletion of the
axion abundance is quite sensitive to initial conditions and
parameters of the theory. This ``chaotic'' behavior likely arises from
the rich dynamics of many coupled oscillators with time-dependent
frequencies of the gauge field $k$-modes, which continually enter and
exit a phase of parametric resonance with the axion. 
This chaotic behavior also appears in various preheating models which
use broad parametric resonance~\cite{Kofman:1997yn}.
Depending on the initial conditions, this sensitivity can
greatly enhance the primordial fluctuations of the axion. This is one
of effects that can lead to a significant enhancement
of the matter power spectrum at small scales. It would be
interesting to study how much of this sensitivity persists once the
effects of the axion inhomogeneities are consistently included.

In our results we ignored the backscattering of dark photons into
axions. This effect is naively very small, as one would conclude by
estimating the $\gamma_D \gamma_D \rightarrow \phi \phi$
rate.  However, due to the very large occupation number of dark
photons the $2 \to 2$ estimate is not reliable. In fact, previous work
which used a similar tachyonic production of gauge fields for
preheating and for generating primordial magnetic fields found that
backscattering effects can lead to quantitative changes in the
results~\cite{Adshead:2015pva,Adshead:2016iae,Peloso:2016gqs}. Our preliminary investigations on
the effects of backscattering indicate that in many cases this effect
creates inhomogeneities in the axion field which can have energies
comparable to the energy in the homogeneous component after the latter
has started efficiently transferring energy to the gauge fields.
However, the energy density in this inhomogeneous component of the
axion is still negligible compared to the energy in the gauge 
fields~\footnote{
  After the first version of this work appeared on the Arxiv, 
  Ref.~\cite{Kitajima:2017peg} presented the results of their
  numerical investigation
  which includes the backscattering effects of the gauge fields into
  axion perturbations. They claim that the energy in the axion
  perturbations can become comparable to the energy in the gauge
  fields, which reduces significantly the dilution of the energy in
  the axion condensate. Details of their calculation are relegated to
  a forthcoming publication. If true, this will significantly modify
  our results for higher values of $f_a$. This
  is an important question which requires further investigation,
  especially in light of a contrasting study by
  Ref.~\cite{Adshead:2015pva} which studied the same effect in the
  context of preheating. 
}.
Therefore, even though  a full treatment that includes
the effects of these perturbations might lead to quantitative changes
in the final results, the qualitative picture of very large energy
transfer from the axion condensate to the gauge fields is still
expected to hold. A complete treatment of this effect could reveal
very interesting features for axion DM. In particular, it might show
that at small scales (comparable to the axion Compton wavelength) the
dark matter density perturbations are much larger than the ones seeded
by inflation, and thus lead to axion miniclusters, which are usually
only associated with axion models with PQ symmetry breaking after
inflation~\cite{Hogan:1988mp,Kolb:1994fi}.

\section{Future directions and conclusions}
\label{sec:conclusions}
The QCD axion is a well-motivated dark matter candidate, and there is
a large experimental program for its search that is underway. There
has been relatively little work on simple extensions of the QCD
axion which can dramatically alter the cosmological evolution of the axion. We
present one such extension where the axion has a coupling to a dark
photon. 
This leads to an efficient depletion of the axion density into
dark photons, such that 
large-$f$ QCD axions can be
misaligned by a generic amount from their minimum. This makes the
theoretically motivated large-$f$ parameter space viable.

For a region of parameter space, this additional energy density in
dark photons can be seen in future CMB experiments. Given the QCD
axion mass, an order of magnitude estimate of this energy density can
be obtained (the prediction is uncertain due to the initial
misalignment angle), and hence constitutes a test for this mechanism.
This mechanism also relaxes the isocurvature bounds on axion DM. If
future experiments measure the energy
scale of inflation, they can rule out a large range of axion decay
constants in anthropic models which would still be allowed in our scenario.

We have presented a minimal extension of the QCD axion model which
focused only on the interactions of a dark photon with the axion.
There are many extensions of this simple model, which include
interactions of the dark photon with other SM fields that can lead to
interesting predictions and are worth exploring. There might also be
interactions of the axion with fields other than a dark photon that
may give us a similar energy dissipation mechanism.


\acknowledgments{
  We thank 
Anson Hook, 
Curly Huang,
Lisa Randall, 
Matt Reece, Ben Safdi and Jesse Thaler for discussions and for useful
comments on our manuscript.
We would like to thank 
Francis-Yan Cyr-Racine, 
Peter Graham,
Mehrdad Mirbabayi, 
Konrad Tywoniuk,
Yue Zhao
for useful discussions.
The work of PA is supported by the National Science Foundation (NSF) grants PHY-0855591 and PHY-1216270.
The work of GMT is supported by the U.S. Department of Energy (DOE) grant DE-SC0012012 and 
the NSF Grant 1316699.
The work of WX is supported by DOE under grant contract numbers DE-SC-00012567 and DE-SC-00015476.
}
\appendix

\section{UV completions of the alignment model}
\label{app:uvmodel}
In this appendix we complete the model presented in
\cref{sec:models}, focusing on generating the alignment potential
term,
\begin{align}
  V(a,b)
  &=
  -\Lambda^4 \cos \left(\frac{n a +b}{f} \right)
  \,.
\end{align}
The purpose of this is to show the origin of large $n$ in the coupling. 
In this section, we will only focus on generating this potential
term for the two axions. We do not explicitly write down the 
couplings of the axions to QCD/dark photons for clarity of
presentation, but these can be easily included.

We note that the examples presented below have a large degree of
overlap -- we can use 4D confining dynamics in a 5D axion model, and
all variants of the mechanisms can be ``iterated'' in a clockwork.
\subsection{A 5D model}
We consider a model where axions arise from gauge fields in
5D~\cite{ArkaniHamed:2003wu,Choi:2003wr}. The 4D axions are identified with phases of Wilson loops around
the extra dimension.
The 5D model consists of two gauge fields in 5D, $(A_M,
B_M)$. 
The fifth dimension is compactified on an orbifold $S_1/Z_2$ at a high
scale, $\frac{1}{2\pi R}
\equiv \Lambda$, with boundary conditions such that $(A_5, B_5)$ zero
modes survive at low
energies~\cite{Bai:2014coa,delaFuente:2014aca,Kaplan:2015fuy}. These
serve as the axions $(a,b)$. There is an
additional fermion $\psi$ in 5D which has charges $(n,1)$ under the two
gauge groups. 

\begin{align}
  \mathcal{L}
  &=
  \int d y 
  \sqrt{G}\left[
    -\frac{1}{4g^2} F_{A,\mu\nu} F_A^{\mu\nu}
    -\frac{1}{4g^2} F_{B,\mu\nu} F_B^{\mu\nu}
    - \bar{\psi} (i\slashed{D} +m) \psi
  \right]
\end{align}
Choosing different 5D gauge couplings for the two gauge groups 
does not by itself lead to a hierarchy in
axion couplings, so for simplicity 
we have chosen them to be identical.
If the mass of the fermion is smaller than the 5D Kaluza-Klein scale,
it generates a sizeable Casimir potential for $(A_5, B_5)$. The
4D effective Lagrangian is
\begin{align}
  \mathcal{L}
  &=
  2\pi R
  \left[
    \frac{1}{2g^2}\partial_\mu A_5 \partial^\mu A_5 
    +\frac{1}{2g^2} \partial_\mu B_5 \partial^\mu B_5
    +\frac{3}{4\pi^2(2\pi R)^5} \cos (2\pi R (n A_5 + B_5))
  \right]
\end{align}
We identify $\sqrt{2\pi R g^2} = 1/f$, such that the Lagrangian for
canonically normalized fields $(a,b)$ is,
\begin{align}
  \mathcal{L}
  &=
  \frac{1}{2}\partial_\mu a \,\partial^\mu a 
  +\frac{1}{2}\partial_\mu b \,\partial^\mu b 
  +
  \frac{3\Lambda^4}{4\pi^2} \cos \left(\frac{n a + b}{f}\right)
  \,.
\end{align}
We have neglected higher harmonics of the cosine potential and finite
$\psi$-mass effects which are subdominant to this term, and do not
affect the alignment qualitatively. 
We note that the coupling of the axions to QCD and the dark photon can
be obtained in this model by including appropriate mixed
Chern-Simons terms.

\subsection{Confining dynamics}
A mechanism along the lines of above can be obtained by a non-abelian
gauge group $SU(N)$ that confines at a high scale $\Lambda$.
We assume $n$ flavors of fermions in the fundamental representation of
$SU(N)$ that are charged under the $U(1)_{PQ,a}$
and a single $SU(N)$ fundamental that is charged under the $U(1)_{PQ,b}$.
\begin{align}
  \begin{array}{c|ccc}
    & SU(N) & U(1)_{PQ,a} & \\
    \hline
    Q &  N  & 1 & \multirow{2}{*}{{\Large\}}$\times n$}\\
    Q^c & \bar{N} & 1 & \\
   \Phi_a & 1 & 2 &
  \end{array}
  &\qquad\qquad
  \begin{array}{c|cc}
    & SU(N) & U(1)_{PQ,b} \\
    \hline
    q &  N & 1\\
    q^c & \bar{N} & 1 \\
   \Phi_b & 0 & 2
  \end{array}
\end{align}
This is very similar to the KSVZ~\cite{Kim:1979if,Shifman:1979if} model. When
the $U(1)_{PQ}$'s are spontaneously broken at scale $f$, the fermions
$Q,q$ get masses $\sim f$. The phases of $\Phi_{a,b}$ couple to the
$SU(N)$
field strength with the following anomaly coefficient,
\begin{align}
  \mathcal{L}
  &=
  \frac{\alpha}{2\pi} \left(\frac{n a}{f} + \frac{b}{f}\right) G \tilde{G}
\end{align}
where $\alpha$ and $G$ are the coupling and field strength for $SU(N)$
respectively.
This results in the axion potential
\begin{align}
  V(a,b)
  =
  -\Lambda_{SU(N)}^4 \cos \left(\frac{n a +b}{f} \right)
  \,.
\end{align}
We now demonstrate that we can obtain a large confinement scale for
the $SU(N)$ gauge
group such that one mass eigenstate is heavy and decouples from the
low-energy theory.  We assume that the $SU(N)$ gauge theory is weakly
coupled at the scale $f$, such that the 't Hooft coupling $\lambda(f)
= 4\pi \alpha(f) N \lesssim 1$. We also choose the $SU(N)$ theory to
be asymptotically free, 
\begin{align}
  b_+  = -\frac{11}{3} N +\frac{2}{3}(n+1) < 0
\end{align}
such that the theory remains weakly coupled in the UV.
Here $b_+$ is the coefficient of the $\beta$-function for the gauge
coupling above the scale $f$.
The dimensionally transmuted scale is,
\begin{align}
  \Lambda_{SU(N)} 
  &=
  f \exp\left(\frac{2\pi }{\alpha(f)\, b_- }\right)
\end{align}
where $b_- = -\frac{11}{3} N$ is the coefficient of the IR $\beta$
function. We see that we can get a high confinement scale, which
decouples the heavy axion. 

Note that there are a large number of degrees of freedom in this
model, $\mathcal{O}(n^2)$, which can renormalize the
Planck scale to be too low. To ameliorate this, we can use the
clockwork mechanism outlined below.

\subsection{Clockwork}
The models considered above can be extended to include more axions,
such that the large charges encountered in those models are
traded for a specific charge
structure~\cite{Dvali:2007hz,Kaplan:2015fuy,Giudice:2016yja,Farina:2016tgd,Craig:2017cda}.

Consider a model with $m+2$ scalars each of which
get a vev $f$ which spontaneously breaks a $U(1)$, resulting in $m+2$
axions. We add the following interaction between these scalars,
\begin{align}
  \mathcal{L}
  &=
  \kappa (\Phi_b^\dagger \phi_1^q + \phi_1^\dagger \phi_2^q+ 
  \ldots + \phi_m^\dagger \Phi_a^q)
  +h.c.
\end{align}
These couplings explicitly break $U(1)^{m+2} \to
U(1)$. The would-be Goldstones from spontaneous breaking of these
symmetries are then massive and can be decoupled from the low-energy
theory. The mass matrix for the Goldstones is 
\begin{align}
  \mathcal{M}
  &=
  \kappa
  f^{q+1}
  \left[
  (b- q\pi_1)^2
  + (\pi_m-q a)^2
  + \sum_{i=1}^{m-1}(\pi_i - q \pi_{i+1})^2
\right]
\end{align}
We have not included the contribution of QCD to the Goldstone masses,
so there is one eigenstate of $\mathcal{M}$ with a zero eigenvalue.
We are interested in the projection of the $a,b$ fields on to this
eigenstate, $\phi$. 
It is easily seen that the normalized eigenvector with the zero
eigenvalue is $\frac{1}{\mathcal{N}}\{1,\frac{1}{q},\ldots
\frac{1}{q^{m+1}}\}$, with 
\begin{align}
\mathcal{N} &= \sqrt{\frac{1-q^{-2(m+1)}}{1-q^{-2}}}
\end{align}
which is $\mathcal{O}(1)$ for $q,m > 1$.
Thus, 
\begin{align}
  b &= \frac{1}{\mathcal{N}}\phi + \ldots \\
  a &=\frac{1}{\mathcal{N}} \frac{1}{q^{m+1}} \phi + \ldots
  \end{align}
Therefore we see that the
effective hierarchy in charges that we get is $n =
q^{m+1}$. For $q=3, m=4$, we easily get $n=\mathcal{O}(100)$.

\bibliographystyle{utphys}
\bibliography{ref}
\end{document}